\documentclass[preprint,prl,citeautoscript]{revtex4}

\usepackage{amssymb}
\usepackage{amsmath}
\usepackage{amsfonts}

\usepackage{graphicx}

\newcommand{\lbrs}{\left[}
\newcommand{\rbrs}{\right]}

\renewcommand{\t}[1]{\mbox{\boldmath$#1$}}
\renewcommand{\d}{{\text{d}}}

\newcommand{\D}[1]{\partial_{#1}}

\newcommand{\Tr}[1]{{{\text{Tr}}\lbrs #1 \rbrs}}
\newcommand{\half}{\frac{1}{2}}
\newcommand{\aeq}{\approx}

\newcommand{\vdW}{\text{vdW}}

\newcommand{\eV}{{\text{eV}}}

\newcommand{\Ang}{\text{\AA}}

\newcommand{\el}{{e^{-}}}

\newcommand{\eref}[1]{equation (\ref{#1})}

\newcommand{\erefs}[2]{equations (\ref{#1}-\ref{#2})}
\newcommand{\rcite}[1]{ref.\ [\onlinecite{#1}]}
\newcommand{\rcites}[1]{refs.\ [\onlinecite{#1}]}

\newcommand{\dn}{\delta n}
\newcommand{\dv}{\delta v}
\newcommand{\tchi}{{\t{\chi}}}
\newcommand{\chib}{{\bar{\chi}}}
\newcommand{\wb}{{\bar{w}}}
\newcommand{\cwb}{{\mathcal{C}}}

\newcommand{\FF}{\mathcal{F}}
\newcommand{\FG}{\mathcal{G}}
\newcommand{\GFN}{\mathcal{M}}

\newcommand{\pin}{p^{-1}}
\newcommand{\zp}{{z^{\prime}}}

\newcommand{\omegaptw}{\omega_{\rm{p2D}}}
\newcommand{\omegapth}{\omega_{\rm{p3D}}}

\newcommand{\vq}{\mathbf{q}}
\newcommand{\vqp}{\vq^{\prime}}

\newcommand{\comment}[1]{}
\newcommand{\ssection}[1]{} 

\newcommand{\new}[1]{#1}

\begin{document}
\title{van der Waals dispersion power laws for cleavage, exfoliation %
and stretching in multi-scale, layered systems}
\author{Tim Gould}
\email{t.gould@griffith.edu.au}
\author{Evan Gray}
\author{John F. Dobson}
\affiliation{Nanoscale Science and Technology Centre, Nathan campus,
Griffith University, 170 Kessels Road,  Nathan, QLD 4111, Australia}
\affiliation{CSIRO National Hydrogen Materials Alliance,
CSIRO Energy Centre, 10 Murray Dwyer Circuit, Steel River Estate,
Mayfield West, NSW 2304, Australia}
\begin{abstract}
Layered and nanotubular systems that are metallic or graphitic
are known to exhibit unusual dispersive
van der Waals (vdW) power laws under some circumstances.
In this paper we investigate the vdW power laws of
bulk and finite layered systems and their interactions
with other layered systems and atoms in the electromagnetically
non-retarded case.
The investigation reveals substantial difference between
`cleavage' and `exfoliation'
of graphite and metals where cleavage obeys a $C_2 D^{-2}$ vdW power law
while exfoliation obeys a $C_3 \log(D/D_0) D^{-3}$ law for
graphitics and a $C_{5/2} D^{-5/2}$ law for layered metals.
This leads to questions of relevance in the interpretation of experimental
results for these systems which have previously assumed more
trival differences.
Furthermore we gather further insight into the effect of scale
on the vdW power laws of systems that simultaneously exhibit
macroscopic and nanoscopic dimensions.
We show that, for metallic and graphitic layered systems,
the known ``unusual'' power laws can be reduced to
standard or near standard power laws when the effective scale of
one or more dimension is changed. This allows better
identification of the systems for which the commonly employed
`sum of $C_6 D^{-6}$' type vdW methods might be valid such as
layered bulk to layered bulk and layered bulk to atom.
\end{abstract}

\maketitle

Layered bulk systems such as graphite and boron nitride have
their atoms confined to a series of
spatially discrete planes with interplanar distances significantly
greater than the intraplanar atomic separations eg. 3.34\Ang\ vs
1.41\Ang\ for graphite.
They can exhibit unusual electronic behaviour
due to the nanometer scale of the layer thickness and macroscopic
scale of the other two dimensions.
This scale variation means that great care must be taken in
investigating subtle physical effects such as dispersion forces as previous
examples demonstrate \cite{Dobson2006,White2008,Gould2008}.

All separated electronic systems exhibit long-range attractive
potentials arising from instantenous electron fluctuations
correlating via the Coloumb potential. These long-range
potentials (often called van der Waals dispersion potentials
when electromagnetic retardation is ignored) are typically
absent from the commonest \emph{ab initio} calculations such as DFT in
the LDA or GGA, or are approximated by
pair-wise inter-atomic potentials of the form $C_6 D^{-6}$ which
are `summed over' in some way (see eg. refs
\cite{Hasegawa2004,Girifalco2002,Rydberg2003,Dappe2006,Ortmann2006,Hasegawa2007,Rydberg2003})
to obtain a new effective power law of the form $C_n D^{-n}$. Here the exponent
$n$ is an integer and depends only on the geometry of the system
while $C_n$ depends on the individual atoms as well as the geometry.

As summarised in \cite{Dobson2006} the exponent of an
asymptotic power law in metallic and graphitic systems can
depend on both the geometry and the type of material, with metals,
graphene and insulators all differing.
For example, with two parallel, nano-thin layers of a metal, graphene and
insulator, the power-law exponents are $5/2$, $3$ and $4$ respectively
(where insulators do obey a `sum over $D^{-6}$' rule).
When the number of layers is infinite
we will show that the power depends not only on the type of material,
but also the way the layers are divided. This will be investigated through
three types of division: equal separation of all layers (`stretching'),
division into two sub-bulks (`cleavage') and removal of one layer from
the top of a bulk (`exfoliation'). Furthermore the interaction of
layered bulks with atoms will be studied.

In this paper we investigate metallic and graphitic systems
under these different types of division
(insulators have trivial `sum over atoms' exponents and need no
further investigation).
Neither metals nor graphitic systems are guaranteed to obey
`sum over layer' power laws and special care must be taken to
evaluate their long-range correlation effects. As with previous
work\cite{DobsonChap1994,Pitarke1998,Dobson1999,
Furche2001,Fuchs2002,Miyake2002,GouldThesis,Jung2004,Marini2006,Gould2008}
we make use of the Adiabatic-Connection Formula and
Fluctuation Dissipation Theorem (ACFFDT) under the
random-phase approximation (RPA) to calculate the leading
power laws under these different methods of division.
All results in this paper are for the electromagnetically
non-retarded case \new{which \cite{Sernelius1998}
show to be unimportant in the range of interest for similar systems}.

Stretched graphitic systems (`straphite') have already been studied
in \cite{Gould2008} where it was shown that the dispersion potential
for an infinite number of graphene layers, each with
inter-layer spacing $D$ follows a $C_3 D^{-3}$
asymptotic power law at $T=0$K where $C_3=0.80 \eV\Ang^3$.
We may make use of the same basic approach employed for
graphitic layered systems to calculate the dispersion for
metallic layered systems (we believe that graphite-metal
intercalates may be examples of this type of system).
For brevity we define an intralayer Coloumb potential
multiplied by a density-density response function
\begin{align}
\cwb(q,u)=\chib(q,u)\wb(q)
\label{eqn:cwb}
\end{align}
where $q$ is the wavenumber parallel to the plane,
$\chib(q,u)$ is the non-interacting electron density-density
response in a layer
and $\wb(q)$ is the Coloumb potential.
When the system is metallic this
takes the form $\cwb(q,u)=-\omegaptw(q)^2 /u^2$ for $q \to 0$
where $\omegaptw(q)=(\frac{n_0 e^2 q}{2 \epsilon_0 m_e})^{1/2}$,
$n_0=N_{\el\rm{layer}}/A_{\rm{layer}}$ is the 2D electron
density of each layer and $m_e$ is the mass of an electron.
We may thus utilise equations 4 and 7 of \rcite{Gould2008} and
change variables to $\theta=qD$ and $\eta=\sqrt{D}\omegaptw(q)^{-1} u$
to show that the difference in the correlation energy per
layer from the infinitely separated case takes the form
\begin{align}
U_{\vdW}\sim \frac{1}{D^{5/2}} \int_0^1 \d\lambda \int_0^{\infty}
\theta^{3/2}\d\theta \int_0^{\infty} d\eta
\FG(\theta,\eta,\lambda)
\end{align}
demonstrating a $C_{5/2}D^{-5/2}$ dispersion power law for layered metals.
Numerical evaluation gives $C_{5/2} = 9.26 \hbar
(\frac{n_0 e^2}{2 \epsilon_0 m_e})^{1/2}$.
As with graphitic systems this power law (although not the constant
prefactor) is universal to all multi-layered metallic systems with
an infinite stack of isotropically stretched layers.

\begin{table}
\begin{tabular}{p{2.8cm}p{1.8cm}p{1.8cm}p{0.8cm}}
System & $C_n^{\rm{system}}/C_n^{\rm{bi}}$ & Predicted & Error
\\
\hline
Tri-graphene & 1.3632 & 1.4167 & 3.6\%
\\
Stretched graphite & 2.1157 & 2.4041 & 14\%
\\
Tri-metal & 1.3496 & 1.4512 & 7.5\%
\\
Stretched metal & 2.0628 & 2.6830 & 30\%
\\
\hline
\end{tabular}
\caption{Comparison of the system to bi-layer ratio
of the per-layer vdW coefficients for
graphitic and metallic systems, for tri-layered and stretched systems.
Column 3 is obtained by summing over the individual layers
and takes the form $\frac{2}{3}(2 + 2^{-n})$
for tri/bi and $2\zeta(p)$ for stretched/bi.}
\label{tab:CoeffComp}
\end{table}
Given the universality of the van der Waals exponent
for isotropic stretching
of a given material it is worth exploring the
validity of a `sum over $C_nD^{-n}$' rule for layered systems.
In Table \ref{tab:CoeffComp} we present the ratio of the $C_n$
coefficient for a tri-layered or stretched system to the
bi-layered system for metals and straphite as well as a
the `sum over $C_n$' prediction for this ratio (here $n$ is $3$
for straphite and $5/2$ for metals). The prediction is somewhat
sound for straphite with an overprediction of $14\%$ for straphite but
is much less so for metals where it leads to a $30\%$ overprediction
of the potential. This suggests that, even using a correct power law,
rules which effectively sum the coefficients may prove troublesome.

\ssection{Cleavage of two bulks}

`Cleavage' represents another means of division of a layered system.
Here the system is split between a single pair of layers to
form two new layered systems. We refer to the new systems as
half-bulks as opposed to the original full bulk. For
homogenous, infinite layered systems the two are mirror
images of each other.

\begin{figure}
\includegraphics[width=70mm]{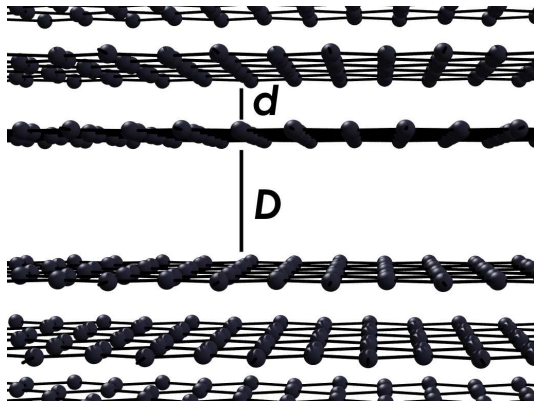}
\caption{Layers after `cleavage'}
\label{fig:Cleaved}
\end{figure}

Separating a layered bulk into two smaller half-bulks
is equivalent to keeping
all but one layer at an inter-layer spacing $d$ while increasing the
remaining one to $D\gg d$ as in Figure \ref{fig:Cleaved} (we ignore
any relaxation of the layers or layer spacing at the newly created
surfaces).
Here we are interested in the
total dispersion energy per unit area rather than that per layer
per area as in the previous case.

We may make use of the second-order perturbation formula
for the dispersion energy between the half-bulks while treating
that within them to all orders (sometimes called the Zaremba-Kohn
\cite{Zaremba1976}
formula). Thus
\begin{align}
U_{\vdW}=\frac{-\hbar}{4\pi^2}\int_0^{\infty} \d u \int_0^{\infty} q \d q
\FF(q,u;D)
\label{eqn:ZK1}
\end{align}
where
\begin{align}
\FF(q,u;D)\aeq 
e^{-2qD} \Tr{ (\tchi_{\lambda=1}(q,u;d)\t{w}^b(qD))^2 }
\label{eqn:ZK2}
\end{align}
where $w^b_{ij}=\wb e^{-qd(i+j)}$ governs the interaction
only \emph{between} the two separate half-bulks while $\chi_{ij}$
is the full interacting response of the density in layer $i$ to a
potential perturbation at layer $j$ \emph{within} a single half-bulk.

Expanding the trace gives
\begin{align}
\FF = e^{-2qD}
\sum_{ijkl} \chi_{ij} w^b_{jk} \chi_{kl} w^b_{li} =
e^{-2qD} (\sum_i \dn_i p^i)^2
\label{eqn:FTwo}
\end{align}
where $\dn_i$ is the interacting electron density response of
layer $i$ in the half-bulk to an external potential perturbation
$\dv_i=\wb p^i$ with $p=e^{-qd}$.

We may calculate $\dn_i$ through the RPA equation
$\dn_i = \cwb p^i + \cwb \sum_{j\geq 0} p^{|i-j|} \dn_j$
where $\cwb=\cwb(q,u)$ is as defined in \eref{eqn:cwb}.
This gives the following recursion relationship
\begin{align}
\dn_{i+2} + \dn_{i} &=
\lbrs (p + \pin) - \cwb (p - \pin) \rbrs \dn_{i+1}
\label{eqn:bulkRel}
\end{align}
where $\dn_0 = \cwb (1 + \sum_{j\geq 0} p^j \dn_j )$
\comment{, $\dn_1 = \cwb (p + \pin \sum_{j\geq 0} p^j \dn_j
+ (p-\pin)\dn_0)$}
and we require $\dn_{\infty} < \infty$.
Writing a formal power series generating function
$\GFN(x;\cwb)=\sum_{i \geq 0} \dn_i x^i$ transforms \eref{eqn:bulkRel} into
\begin{align}
&(1+x^2)\GFN(x;\cwb) - \dn_0 - x\dn_1=
\nonumber \\& \hspace{5mm}
x\lbrs (p+\pin) + \cwb(p-\pin) \rbrs (\GFN(x;\cwb)-\dn_0)&
\end{align}
so that (with correct series asymptotics)
\begin{align}
\GFN(x;\cwb)&= \frac{\cwb(p-\pin)}{p(1 - r_- x)(1 - r_+\pin)}
\end{align}
where $r_{\pm}=\zeta \pm \sqrt{\zeta^2 -1}$,
$\zeta=\half(p+\pin)+\half \cwb(p-\pin)$. Noting that
$\sum_i p^i \dn_i=\GFN(p)$ lets us write \eref{eqn:FTwo} as
\begin{align}
\FF(q,u;d,D) = e^{-2qD} \GFN(e^{-qd};\cwb(q,u))^2.
\end{align}

For graphite \cite{Dobson2006} shows that
$\cwb(q,u)=\kappa[1+u^2/(v_0 q)^2]^{-1/2}$
where $v_0=5.0 \times 10^5\rm{ms}^{-1}$,
$\kappa=\frac{e^2}{4\epsilon_0\hbar v_0}=12.1$.
Defining $\alpha=d/D$, $\theta=qD$ and $\eta=u d/v_0$
lets us expand $\FF(\theta,\eta)=\FF(qD,u d/v_0;\alpha,1)$
in powers of $\alpha$. To leading order this gives
\begin{align}
\FF(\theta,\eta)=e^{-2\theta}
\frac{\kappa^2}{(\kappa + \eta + \sqrt{\eta^2 + 2 \kappa \eta})^2}
\end{align}
for graphite. A similar change of variables
yields a similar leading order expansion for metals. For these cases
\eref{eqn:ZK1} gives a leading power law of exponent $n=2$ as
with insulators.

Thus for grapite and metals we can write
\begin{align}
U_{\vdW}(D\to\infty) \aeq \frac{C_2}{D^{2}}
\end{align}
where
\begin{align}
C_2=\frac{-\hbar}{4\pi^2}\frac{\Gamma(2)}{2^2} \int_0^{\infty} \d \eta
T(\eta).
\end{align}
Here $T(\eta)=\frac{v_0}{d} \kappa^2
( \kappa + \eta + \sqrt{\eta^2 + 2 \kappa \eta})^{-2}$ for graphite
and $T(\eta)=\frac{\omegapth}{\sqrt{2}}
(  1 + \eta^2 + \eta\sqrt{\eta^2 + 2})^{-2}$ for metals with
$\omegapth=(\frac{n_0 e^2}{\epsilon_0 m_e d})^{1/2}$.
For graphite we find $C_2=0.13d^{-1} \eV\Ang^{3}$
and for metals $C_2=1.6886 \times 10^{-3} \hbar \omegapth$.
The latter result agrees exactly with continuous
but anisotropic models of half-bulk metals where electron
movement is restricted to be parallel to the surface,
as is expected if the $d\to 0$ limit is well defined.

These results differ significantly from those expected by a
simple sum-of-layers approach where we would expect
graphene to obey a $C_1 D^{-1}$ power law, and metals
to obey a $C_{1/2} D^{-1/2}$ power law. The screening in these
layered systems seems to cancel the different correlation effects
of the individual layers so that they act as pseudo-insulating bulks.

\ssection{Exfoliation}
Another means of dividing graphene (`exfoliation')
is to peel a single layer of graphene from the top of a bulk.
This represents yet another method of division
where one system is a layered half-bulk and the other a single layer.
We restrict our investigation to the case where the removed layer plane
is always stiff and parallel to the planar surface of the bulk.

To model exfoliation we use a similar perturbative approach to
cleavage but in \eref{eqn:ZK1} set $\FF =  e^{-2qD} \GFN(e^{-qd};\cwb)
\frac{\cwb}{1+\cwb}$ the product of the interacting
response of a half-bulk and a single layer.
For graphene this complicates the problem
as the $\alpha=d/D \to 0$ limit works well for $\GFN$ but not for
$\frac{\cwb}{1+\cwb}$. Here we keep $\alpha$ in the single-layer response
\emph{only}. Under transformation of variables $\theta=qD$
and $\eta=\frac{\alpha}{v_0}u$ (which already gives a $D^{-2}$ outside
the integrals) we find
\begin{align}
\FF =&\frac{v_0}{D} \kappa^2 \theta e^{-2\theta}
(\kappa + \eta + \sqrt{\eta^2 + 2 \kappa \eta})^{-1}
\nonumber\\&\hspace{10mm}\times
(\sqrt{\eta^2 + (\alpha\theta)^2} + \kappa \alpha\theta)^{-1}.
\end{align}
Setting $\theta=1$ (where $\theta^2 e^{-2\theta}$
takes its maximum), approximating $\sqrt{\eta^2+\alpha^2}$
by $\alpha+\eta$ and $\kappa+\eta+\sqrt{\eta^2+2\kappa\eta}$ by
$2(\kappa+\eta)$
allows us to approximate the $\eta$ integral to show a leading
$\alpha\log(\alpha)$ term.
This is equivalent to a power law of the form
\begin{align}
U_{\vdW}(D\to\infty) \aeq \frac{C_3 \log(D/D_0)}{D^3}.
\end{align}
and numerical calculation of \eref{eqn:ZK1} validates this
assumption. For graphene (where $\kappa=12.1$) we find
$D_0=0.16 d$ and $C_3= 0.07 \eV\Ang^3$.

A similar analysis of metals shows a $C_{5/2} D^{-5/2}$
power law with $C_{5/2}=6.42 \times 10^{-3} \sqrt{d} \hbar \omegapth
= 9.07 \times 10^{-3} \hbar \omegaptw(q=1)$.
The $5/2$ power law is the same as that of a metal layer interacting
with a continuum model of a  metallic half-bulk as is,
again, predicted by the limit of $d\to 0$.

While this analysis is not valid in the small $D$ regime it is
worth noting that exfoliation and cleavage exhibit different
power laws for graphite. This suggests that sum of $C_6$ models
for converting experimental results from one to the other
such as those employed in \rcite{Benedict1998}
may need reexamination.
Unfortunately accurate calculation of the dispersion
energy of such systems for $D \aeq d_0$ (where $d_0$ is the layer
spacing of graphite) is as yet intractable.

\ssection{Atoms}

In nanoscale systems there are often combinations of molecules,
layers and bulks. A simple example is an atom interacting with
the surface of a layered metal or a molecule interacting with a
graphene surface. Here the power law could be affected by the
layering and electronic properties of the material.

In the coordinate system used for the layered models
the interacting response function of an infinitely small
``atom'' located at $Z\hat{z}$ can be written as
\begin{align}
&\chi_A(\vq,\vqp,z,\zp;u)=
\nonumber \\
&\hspace{5mm} \alpha(iu) \lbrs
\vq \cdot \vqp
+ \D{z}\D{\zp}
\rbrs \delta(z-Z) \delta(\zp-Z)
\label{eqn:chiatom}
\end{align}
where $\vq$ and $\vqp$ are reciprocal lattice vectors in the plane and
$\alpha(iu)$ is the interacting dipole polarisibility of the
atom at imaginary frequency $u$. This formula, used in
\erefs{eqn:ZK1}{eqn:ZK2}, correctly reproduces
the $C_6 D^{-6}$ power law for interacting atoms.

We can use \eref{eqn:chiatom} to calculate the
interaction of an atom with a layered bulk (metallic or graphitic) by
making use of \erefs{eqn:ZK1}{eqn:ZK2}.
This gives a power-law exponent $n=3$
in agreement with the prediction of a sum over $C_6 D^{-6}$ potentials.
This result strongly suggests that the unusual power laws
exhibited by layered systems result from the interaction
between long-range fluctuations in \emph{both} systems
and that removing them from one reduces the systems to `typical'
dispersive behaviour.

\ssection{Results}

As has been seen here and in other work\cite{Dobson2006,White2008,Gould2008}
the asymptotic power law behaviour of layered systems can be anything
but simple.
Both graphitic and metallic systems exhibit vastly different dispersive
power laws to insulators so that `sum of $C_6$' approximations
such as those typically employed cannot
be used in the asymptotic region. It seems unlikely that, with such
varied asymptotes, the cohesive energies and other similar
measurables can be investigated using simple models.

\begin{table}
\begin{tabular}{lp{2cm}p{2cm}p{2cm}}
 & Graphite & Metal & Insulator
\\
\hline
Stretching & $D^{-3}$ & $D^{-5/2}$ & $D^{-4}$
\\
Cleavage & $D^{-2}$ & $D^{-2}$ & $D^{-2}$
\\
Exfoliation & $\log(\frac{D}{D_0}) D^{-3}$ & $D^{-5/2}$ & $D^{-3}$
\\
Atom-bulk & $D^{-3}$ & $D^{-3}$ & $D^{-3}$
\end{tabular}
\caption{Asymptotic power laws for various systems demonstrating both
the material and structural dependence.}
\label{tab:Powers}
\end{table}
To illustrate these discrepancies we present in Table \ref{tab:Powers}
a summary of the various power laws studied here.
The insulator result represents the
`classic' sum over atomic power-laws behaviour of each system and any
difference from its exponent represents `unusual' behaviour.

While uniform stretching results in different power laws for metals,
graphitics and insulators, cleavage removes this variation and
involves the same exponent for all materials.
This suggests that the interlayer screening
induced by the Coulomb potential dominates the local response of each
layer in the van der Waals energy for such systems converting their
behaviour into that of non-layered or insulating bulks. This is
further demonstrated by the fact that both the power law
\emph{and} coefficient of cleaved layered metals is the
same as that of cleaved bulk metals with electron movement restricted
to the plane.

By contrast, keeping a finite number of layers asymptotically isolated,
as in exfoliation, or all layers asymptotically separated, as in
stretching, returns different power-laws
for different systems. In these cases at least one layer can be
considered infinitesimally thin which we believe to be a requirement
for the unusual power-laws.
Replacing the isolated layer by an atom, however, returns the
classical results which suggests that at least one large dimension
is required for unusual vdW dispersion power-laws as postulated in
\rcites{Dobson2006,White2008}

Overall, as this work and references \cite{Dobson2006,White2008,Gould2008}
demonstrate, the dispersion forces of systems with a mix of nanometre and
macroscopic length scales are more complex than classic Lifshitz
theory predicts.
As the differing power laws for cleavage and exfoliation demonstrate
we must take great care in using indirectly derived
cohesive energies from experiment.

These unusual van der Waals power laws may also have profound effects
on the behaviour of many nanosystems. For certain systems it may
be neccessary to adapt molecular dynamics and other
semi-empirical and approximate
\emph{ab initio} simulation methods to account for these differences
in order to best replicate experiment.

\ssection{Acknowledgements}
The authors acknowledge funding from the NHMA.

\end{document}